\documentclass{PoS}
\pdfoutput=1

\usepackage{amsmath}
\usepackage{bbold}
\usepackage{pgfplots}
\usepackage{graphicx}

\definecolor{myblue}{RGB}{35,55,145}
\definecolor{myred}{RGB}{220,25,25}

\DeclareMathOperator{\Tr}{Tr}
\DeclareMathOperator{\SU}{SU}
\DeclareMathOperator{\U}{U}
\DeclareMathOperator{\su}{su}

 \newcommand{\la}{\left\langle}
 \newcommand{\ra}{\right\rangle}
 \newcommand{\gi}{\tilde g_I}

\title{Induced YM theory with auxiliary bosons\thanks{Work supported by DFG in the framework of SFB/TRR-55. We thank Christoph Lehner for discussions at an early stage of this project.}}

\ShortTitle{Induced YM theory with auxiliary bosons}

\author{Bastian B. Brandt, \speaker{Robert Lohmayer}, and Tilo Wettig\\
Department of Physics, University of Regensburg, 93040 Regensburg, Germany \\
E-mail: \email{robert.lohmayer@ur.de}}

\abstract{
We study pure $\SU(N_c)$ lattice gauge theory with a plaquette weight factor given by an inverse determinant which can be written as an integral over auxiliary bosonic fields (modifying a proposal of Budczies and Zirnbauer). We derive conditions for the existence of a continuum limit and its equivalence to Yang-Mills theory. Furthermore, we perturbatively compute the relation between the coupling constants of the `induced' gauge action and the familiar Wilson gauge action using the background-field technique. The perturbative relation agrees well with numerical results for $N_c=2$ in three dimensions.
}

\FullConference{The 33rd International Symposium on Lattice Field Theory\\
14 -18 July 2015\\
Kobe International Conference Center, Kobe, Japan}

\begin{document}

\section{Introduction}

In general, persistent challenges for lattice QCD, e.g., the sign problem for real chemical potential, critical slowing down, or volume scaling, make it worthwhile to explore alternative discretizations of Yang-Mills theory. 
The idea we pursue here is to induce gauge dynamics by auxiliary fields coupled linearly to the gauge fields, rendering applicable analytic methods and simulation algorithms used in strong-coupling expansions. 
While earlier approaches based on that idea required an infinite number of infinitely heavy auxiliary fields (or did not exhibit the desired YM continuum limit), Budczies and Zirnbauer (BZ) succeeded in Ref.~\cite{Budczies:2003za} to induce $\U(N_c)$ pure YM theory with $N_b\geq N_c$ auxiliary fields.  
A slightly modified version of the BZ approach (curing a trivial sign problem) adapted to $\SU(N_c)$ was introduced and investigated numerically in Ref.~\cite{Brandt:2014rca}. Numerical results for $N_c=2$ in three dimensions showed good agreement with simulations using the usual Wilson action.
In the following, we show analytically that the induced $\SU(N_c)$ gauge action indeed exhibits a continuum limit (for $N_b\geq N_c-\frac54$) which is equivalent to YM theory for $N_b\geq N_c-\frac34$ in two dimensions. Universality arguments \cite{Budczies:2003za} as well as numerical results \cite{Brandt:2014rca} suggest that this equivalence persists in higher dimensions. 
Furthermore, we perturbatively compute the relation between the coupling constants of the induced gauge action and Wilson's gauge action using the background-field technique. For technical details, we refer to Ref.~\cite{InducedDraft}.

\section{Induced lattice gauge action}

Instead of the familiar Wilson weight factor $\omega_W(U_p)=\exp\left[\frac1{g_W^2}\Tr\left(U_p+U_p^\dagger-2\right)\right]$ for the plaquette variables 
\begin{align}%
U_p \equiv U_\mu(x) U_\nu(x+\hat\mu)U_\mu^\dagger(x+\hat\nu)U_\nu^\dagger(x) \in\SU(N_c)\,,\qquad
p\equiv(x;\mu<\nu)\,,
\end{align}%
we consider the pure gauge plaquette weight factor \cite{Brandt:2014rca}
\begin{align}\label{Eq:omega}
\omega(U_p)&={\det}^{- N_b}\left(1-\frac \alpha2 (U_p+U_p^\dagger)\right)
\end{align}
with $0<\alpha\leq 1$ and $N_b>0$ not necessarily restricted to integer values.

For integer $N_b$, however, the inverse determinants can be written as integrals over $N_b$ complex auxiliary boson fields (in the fundamental representation) with mass parameter $m\geq 2$ determined by the parameter $\alpha$ through $\frac 2\alpha=m^4-4 m^2+2$,
\begin{align}%
\prod_p \omega(U_p)&=\int \left[d\phi\right] e^{- S_B[\bar \phi, \phi, U]}\,, \\
S_B [\bar \phi,\phi,U] &= \sum_{b=1}^{N_b} \sum_{p} \sum_{j=1}^{4}
  \Big[ m \bar \phi_{b,p}(x^p_j)\phi_{b,p}(x^p_j) - \bar \phi_{b,p}(x^p_{j+1})
  U(x^p_{j+1},x^p_j)\phi_{b,p}(x^p_j) \Big. \notag\\
  &\qquad\qquad\qquad- \bar \phi_{b,p}(x^p_{j}) U(x^p_j,x^p_{j+1}) \phi_{b,p}(x^p_{j+1})\Big]\,,
\end{align}%
where  $U(x+\hat \mu,x)\equiv U_\mu(x)$ etc.~and $x^p_j$ with $j=1,\ldots,4$ denote the four points of the plaquette $p$ in the order of appearance following the plaquette orientation.

\section{Continuum limit}

In order to prove that the induced lattice gauge action indeed exhibits a continuum limit, we determine whether $\omega(U_p)$, as given in Eq.~\eqref{Eq:omega}, reduces to a $\delta$-function (located at $U_p=\mathbb{1}$) on the $\SU(N_c)$ group manifold in the limit $\alpha \to 1$. To this end, we make use of the Peter-Weyl theorem,
\begin{align}\label{Eq:Peter-Weyl}
\delta(U) \propto \sum_{\text{all irreps}\ \lambda} d_\lambda \chi_\lambda(U)\,,
\end{align}
where the sum is over all irreducible representations $\lambda$ with dimension $d_\lambda$ and character $\chi_\lambda$. 
Similarly expanding $\omega(U)$ in irreducible characters,
\begin{align}\label{Eq:clambda}
\omega(U) = \sum_{\lambda} c_{\lambda}(\alpha) \chi_\lambda(U)\,,
\qquad \qquad
c_\lambda(\alpha) = \int dU \omega(U) \chi_\lambda(U^{-1})\,,
\end{align}
we obtain the properly normalized weight function $\bar \omega(U)$ through
($\lambda=0$ for the trivial rep.)
\begin{align}%
 \mathcal Z\equiv\int dU \omega(U) = c_0(\alpha)\,,\qquad\qquad
 \bar \omega(U)\equiv \frac 1{\mathcal Z} \omega(U) = \sum_{\lambda} \frac{c_{\lambda}(\alpha)}{c_0(\alpha)} \chi_\lambda(U)\,.
\end{align}%

 To compute $\lim_{\alpha\to 1} c_\lambda/c_0$, we parametrize the plaquette variable as $U=e^{i \sqrt{\gamma(\alpha)} \, H}$ with $\gamma(\alpha)=\frac 2\alpha(1-\alpha)$ and $H$ restricted to a domain $V(\alpha)\subset \su(N_c)$ such that $\SU(N_c)$ is covered exactly once. The corresponding integration measure is given by 
\begin{align}%
dU   = \gamma(\alpha)^{\frac{N_c^2-1}2} \sqrt{\det g(H)}\,\, dH\,, \qquad
g(H) =\frac12+\sum_{k=1}^\infty \frac{(-1)^k}{(2k+2)!}\gamma(\alpha)^k H^{2k}_\text{adjoint}
\end{align}%
with $H_\text{adjoint}$ denoting the element of the adjoint representation of $\su(N_c)$ corresponding to $H$ in the fundamental representation.
Next, we expand the integrand of Eq.~\eqref{Eq:clambda} in powers of $1-\alpha$, using 
\begin{align}%
{\det}^{-N_b}\left(1-\frac\alpha 2\left(U+U^\dagger\right)\right)
&=
\frac {{{\det}^{-N_b}\left(1+ H^2\right)}}{(1-\alpha)^{N_b N_c}}\left(1+N_b \frac{1-\alpha}{6\alpha} \Tr \frac{H^4}{1+H^2}+ \ldots \right) \,,\\
\chi_\lambda\left(e^{-i \sqrt{\frac2\alpha(1-\alpha)} \, H}\right) &={d_\lambda}\left(1-\frac{ C_2^{\SU(N_c)}(\lambda) }{N_c^2-1} (1-\alpha) \Tr H^2+\ldots\right) \,,
\end{align}%
where $C_2$ denotes the quadratic Casimir invariant of $\SU(N_c)$. 
We then integrate over $V(\alpha)$ term by term and analyze the asymptotic behavior of these integrals in the limit $\alpha \to 1$.

In Ref.~\cite{InducedDraft} we show in detail that the first-order term of the integrand of Eq.~\eqref{Eq:clambda} dominates over all higher-order terms in the limit $\alpha \to 1$ as long as 
 $\int_{V(\alpha)}dH\,{\det}^ {-N_b}(1+H^2)$ is finite or at most logarithmically divergent as $\alpha \to 1$. 
In the eigenvalue parametrization 
\begin{align}%
\int_{V(\alpha)}dH\,{\det}^ {-N_b}(1+H^2) \propto 
\int_{-\pi/\sqrt{\gamma(\alpha)}}^{\pi/\sqrt{\gamma(\alpha)}}
\left( \prod_{j=1}^{N_c} dz_j \right)\delta\left(\sum_{j=1}^{N_c}z_j\right) \left(\prod_{j<k}(z_j-z_k)^2\right)  \prod_{j=1}^{N_c} \left(1+z_j^2\right)^{-N_b}
\end{align}%
we see that this is the case for $N_b\geq N_c -\frac54$. This immediately results in $c_\lambda(\alpha)/c_0(\alpha) = d_\lambda$ at leading order in $1-\alpha$ and consequently, see Eq.~\eqref{Eq:Peter-Weyl},
\begin{align}%
\lim_{\alpha\to1}\bar \omega(U) = \sum_\lambda d_\lambda \chi_\lambda(U) \propto \delta(U)
\qquad\qquad
\text{for} \quad
N_b \geq N_c -\frac 54\,.
\end{align}%
For $N_b < N_c -\frac 54$, all terms in the expansion of the integrand  of Eq.~\eqref{Eq:clambda} result in identical leading-order divergences as $\alpha \to 1$ (when integrated over $V(\alpha)$), which means that $\lim_{\alpha\to 1} c_\lambda/c_0$ is a non-trivial function of $\lambda$ that generically differs from $d_\lambda$.

We therefore conclude that $N_b\geq N_c -\frac54$ is a necessary condition for the existence of the continuum limit. The same method applied to the $\U(N_c)$ case leads to the condition $N_b \geq N_c-\frac 12$. These bounds have been verified through numerical simulations and explicit analytical calculations for $\SU(2)$, see Refs.~\cite{Brandt:2014rca} and \cite{InducedDraft}.

To determine the nature of the continuum limit (provided that it exists), we need to expand the ratio $c_\lambda(\alpha)/c_0(\alpha)$ to linear order in $1-\alpha$. 
We use the same strategy as before, i.e., expand the  integrand of Eq.~\eqref{Eq:clambda} in powers of $1-\alpha$ and term-by-term integrate over $V(\alpha)$. For the next-order term, the relevant integral is
$
\int_{V(\alpha)}dH\,{\det}^ {-N_b}(1+H^2) \Tr H^2
$
which exists for $N_b>N_c-\frac 34$ and is logarithmically divergent for $N_b=N_c-\frac34$. Concerning the dependence on $\lambda$, we hence obtain for $N_b>N_c-\frac 34$, 
\begin{align}%
\frac{c_\lambda(\alpha)}{c_0(\alpha)}=
{d_\lambda} \left(
1 -(1-\alpha) \frac{{C_2^{\SU(N_c)}(\lambda)} }{N_c^2-1}  \frac{\int_{\su(N_c)}dH\,{\det}^ {-N_b}(1+H^2) \Tr H^2}{\int_{\su(N_c)}dH\,{\det}^ {-N_b}(1+H^2)}+\ldots\right)\,.
\end{align}%
For $N_b=N_c-\frac34$, $1-\alpha$ has to be replaced by $(1-\alpha)\log(1-\alpha)$ in the above expansion. 
For $N_b< N_c-\frac 34$, the NLO term generically has a more involved dependence on $\lambda$, not simply given by $C_2(\lambda)$. 
  
Analogously to the Wilson weight factor for small $g^2$, the induced weight function $\bar \omega(U)$ therefore reduces to the heat-kernel weight factor for small $1-\alpha$,
$\frac{c_\lambda(\alpha)}{c_0(\alpha)}= d_\lambda e^{-\tau\, C_2^{\SU(N_c)}(\lambda)}$
with diffusion parameter $\tau\equiv \tau(\alpha,N_b,N_c)$ as long as $N_b\geq N_c-\frac34$. 
In two dimensions, the heat-kernel lattice action is exactly self-reproducing, i.e., the effective action for a doubled lattice cell, obtained by integrating over all internal link variables, has the same functional form as the original plaquette action (Migdal's recursion \cite{Migdal:1975zg}). Hence, taking the continuum limit is trivial in 2d and we conclude that the continuum limit of the induced theory is in the universality class of $\SU(N_c)$ Yang-Mills theory for $N_b\geq N_c-\frac 34$. (For gauge group $\U(N_c)$, we obtain the condition $N_b\geq N_c+\frac 12$.)

Following Ref.~\cite{Budczies:2003za}, we conjecture that the equivalence with Yang-Mills theory persists also in three and four dimensions since the collective nature of the fields gets enhanced (compared to 2d) which should work in favor of universality. First numerical tests in 3d indeed confirm this to be the case \cite{Brandt:2014rca}.

\section{Relation of coupling constants in perturbation theory}

Since the continuum limit is essentially trivial in two dimensions, the relation of $1-\alpha$ and the Wilson coupling constant $g_W^2$ can be obtained simply by matching the character expansions of the plaquette weight functions. In three and four dimensions on the other hand, the continuum limit is more involved and we need perturbation theory to determine the relation of the coupling constants. 

However, in a perturbative expansion in powers of $1-\alpha$ at fixed $N_b$, we encounter two problems: First, the expansion of the logarithm in (ignoring irrelevant constants)
\begin{align}%
S=N_b \sum_p \Tr \log\left(1-\frac \alpha 2\left(U_p+U_p^\dagger\right)\right)
=
N_b \sum_p \Tr \log\left(1-\frac \alpha {2(1-\alpha)}\left(U_p+U_p^\dagger-2\right)\right)
\end{align}%
is justified only if $\left | \frac{\alpha}{1-\alpha}(\cos\varphi-1) \right | \leq 1$ for all possible eigenvalues  $e^{i \varphi}$ of $U_p$, i.e., $\alpha \leq \frac13$.
Second, after expanding the logarithm anyway, we see that a saddle-point analysis of the partition function is not possible since higher orders of $U+U^\dagger -2$ are not suppressed in 
\begin{align}\label{Eq:Sind}
S=-N_b \sum_p \sum_{n=1}^\infty \frac 1 n \left(\frac \alpha {2(1-\alpha)}\right)^{n}
\Tr \left(\left(U_p+U_p^\dagger-2\right)^{{n}}\right)
\end{align}
and we end up with non-Gaussian integrals. 

As a workaround, we will therefore first keep $\alpha \leq \frac13$ fixed and take the limit $N_b\to\infty$, which allows for a systematic saddle-point analysis, and then analytically continue $g_W(\alpha,N_b)$ to small $1-\alpha$. Since the $(n=1)$-term in Eq.~\eqref{Eq:Sind} corresponds to the Wilson gauge action, it is natural to define the coupling constant $\gi$ for the induced theory as
\begin{align}%
\frac{1}{\gi^2}=N_b \frac{\alpha}{2(1-\alpha)}\,.
\end{align}%

In order to compute the relation of the coupling constants $g_W$ and $\gi$ for fixed $\alpha$, 
\begin{align}\label{Eq:gWgI}
\frac{1}{g_W^2}=\frac1{\gi^2}\left(1+c_1(\alpha) \gi^2 + c_2(\alpha) \gi^4+\ldots\right)\,,
\end{align}
it is convenient to use the background-field technique (based on Refs.~\cite{Dashen:1980vm,Hasenfratz:1981tw}). As usual, we parametrize the link variables in terms of quantum fields $q_\mu$ and background fields $A_\mu$ through
\begin{align}%
U_\mu(x)=e^{i a g q_\mu(x)} U^{(0)}_\mu(x) \,,\qquad U^{(0)}_\mu(x) =  e^{i a A_\mu (x)}
\end{align}%
with $g=g_W$ and $g=\gi$, respectively,
and compute the effective actions $\Gamma_{I/W}[A]$,
\begin{align}%
e^{-\Gamma_{I/W}[A]} \propto  \int_{\text{1-PI}}[Dq] e^{-S_{I/W}[A,q]}\,,
\end{align}%
to quadratic order in the background fields. The relation between the coupling constants $\gi$ and $g_W$ is then obtained by requiring $\Gamma_I[A]=\Gamma_W[A]$ in the continuum limit $g\to 0$.

Since we need to expand the gauge action only to quadratic order in $A$, we write
\begin{align}%
S_I = \left. S_W \right|_{g_W=\gi} + \sum_{n=2}^\infty \left(S_I^{(n,0)}+S_I^{(n,1)}+S_I^{(n,2)}+\mathcal O(A^3) \right)\,,
\end{align}%
where $S_I^{(n,k)}$ includes all $\mathcal O(A^k)$ terms resulting from $\Tr\left(\left(U_p+U_p^\dagger -2\right)^n\right)$ in the sum over $n$ in Eq.~\eqref{Eq:Sind}. 
With $q_{\mu\nu}(x)\equiv q_\mu(x)+q_\nu(x+\mu)-q_\mu(x+\nu)-q_\nu(x)$ and $A_{\mu\nu}$ defined analogously, 
we obtain to leading order in the quantum field 
\begin{align}%
S_I^{(n,0)} & = (-1)^{n+1} \frac{a^{2n}\gi^{2n-2}}{(2/\alpha -2)^{n-1}}  \sum_{x,\mu,\nu} \frac{1}{2n} \Tr\left[ q_{\mu\nu}(x)^{2n}+\mathcal O \left(\gi q^{2n+1} \right) \right]\,, \\
S_I^{(n,1)} & = (-1)^{n+1} \frac{a^{2n}\gi^{2n-3}}{(2/\alpha -2)^{n-1}} \sum_{x,\mu,\nu} \Tr\left[ {A_{\mu\nu}(x)} q_{\mu\nu}(x)^{2n-1}+ \mathcal O \left( \gi A q^{2n} \right)\right]\,, \\
S_I^{(n,2)} & = (-1)^{n+1} \frac{a^{2n}\gi^{2n-4}}{(2/\alpha -2)^{n-1}} \sum_{x,\mu,\nu}
\Tr\Biggl[  \sum_{m=0}^{n-2} {A_{\mu\nu}(x)} q_{\mu\nu}(x)^m {A_{\mu\nu}(x)} q_{\mu\nu}(x)^{2n -m -2} \cr
&\qquad\qquad\qquad\qquad\qquad\qquad\quad
+\frac 12 \left({A_{\mu\nu}(x)}q_{\mu\nu}(x)^{n-1}\right)^{2} +
 \mathcal O \left(A^2 \gi q^{2n-1}\right)\Biggr]\,.
\end{align}%
The expansion of the Wilson action (i.e., the $(n=1)$-term in the induced action) as well as the gauge-fixing procedure can be taken over one-to-one from the Wilson case. 

Splitting the action in a `free' action $S_f$ for the quantum field $q$ (i.e., terms of order $q^2A^0$), a `classical' piece $S_{\text{cl}}[A]$ (terms independent of $q$), and `interaction' terms $S_\text{int}$ (all remaining terms), we get
\begin{align}%
e^{-\Gamma[A]} \propto e^{-S_{\text{cl}}[A]} \int_{\text{1-PI}}[Dq] e^{-S_f[q]} \sum_k \frac 1{k!}\left(- S_{\text{int}}[A,q]\right)^k
\propto e^{-S_{\text{cl}}[A]} \sum_k \frac 1{k!}  \la \left(- S_{\text{int}}[A,q]\right)^k \ra_{\text{1-PI}}\,,
\end{align}%
where we omitted integrals over ghost fields since these do not contribute to LO and relevant NLO terms.
Expectation values are taken w.r.t.~the free action $S_f = \frac {a^4}2  \sum_{x,\mu,b} q_\mu^b(x) \square q_\mu^b(x)$ with lattice d'Alembert operator $\square$,
\begin{align}%
\la q_\mu^a(x) q_\nu^b(y) \ra &= \delta_{ab}\delta_{\mu\nu} D(x-y)\,,
\end{align}%
where $D(x)$ denotes the standard lattice propagator for a massless scalar field.

Since $S_I$ includes $S_W$, the corresponding terms cancel in $\Gamma_I-\Gamma_W$ and the one-loop coefficient $c_1(\alpha)$ in Eq.~\eqref{Eq:gWgI} is exclusively determined from
\begin{align}%
\la S_I^{(2,2)} \ra  
 &=
  - \frac{a^4 \alpha}{2(1-\alpha)} \sum_x \sum_{\mu,\nu} A^a_{\mu\nu}(x) A^b_{\mu\nu}(x) 
\la q^c_{\mu\nu}(x) q^d_{\mu\nu}(x) \ra 
\Tr \left[ t_a t_b t_c t_d +\frac 12 t_a t_c t_b t_d \right] \cr
&\overset{a\to 0}{\to} - \frac 4d \left(\frac{2 N_c^2-3}{8 N_c}\right) \frac \alpha{2(1-\alpha)} a^{4-d} \int d^dx \sum_{\mu,\nu} \Tr F_{\mu\nu}(x)^2+\ldots
\end{align}%
with $\SU(N_c)$ generators $t_a$ normalized to $\Tr t_a t_b = \frac 12 \delta_{ab}$. 
In $d$ dimensions, comparison with 
$S_{\text{cl}}[A] = \frac1{2 \gi^2} a^{4-d} \int d^dx \sum_{\mu,\nu} \Tr F_{\mu\nu}(x)^2+\ldots$
leads to 
\begin{align}%
c_1(\alpha)=c_{1,-1} \left( \frac \alpha {2(1-\alpha)} \right)\,,\qquad\qquad
 c_{1,-1}=- \left(\frac{2 N_c^2-3}{N_c d}\right) \,.
\end{align}%
At order $\gi^2$, $\Gamma_I$ contains terms of order $(1-\alpha)^{-2}$ and $(1-\alpha)^{-1}$,
\begin{align}%
c_2(\alpha)&=c_{2,-2} \left( \frac \alpha {2(1-\alpha)} \right)^2 + c_{2,-1} \left( \frac \alpha {2(1-\alpha)} \right)\,.
\end{align}%
However, we are only interested in the coefficient $c_{2,-2}$ (see Eq.~\eqref{eq:continuation} below), which is found to be given by
\begin{align}%
c_{2,-2}&=\frac{N_c^4-3N_c^2+6}{d^2 N_c^2} - \frac{N_c^4-6 N_c^2+18}{2 N_c^2 (d-1)} \left(\frac 3{d^2}-4(4-d) J_d\right)
\end{align}%
with $J_{2}=\frac 1{32}$ and $J_{3}\approx 0.0085535415$ 
from $\la S_I^{(3,2)} - S_I^{(2,0)}S_I^{(2,2)} -\frac12 S_I^{(2,1)}S_I^{(2,1)} \ra$.

If we now rewrite the RHS of Eq.~\eqref{Eq:gWgI} in terms of $\alpha$ and $N_b$, we obtain
\begin{align}\label{eq:continuation}
\frac{1}{g_W^2} 
&
=\frac \alpha {2(1-\alpha)} \underbrace{\left[N_b + c_{1,-1} + c_{2,-2}/N_b + \mathcal O\left(N_b^{-2}\right)\right]}_{\equiv d_0(N_b)}+\mathcal O((1-\alpha)^0)\,.
\end{align}
For the limit $\alpha\to 1$ at fixed $N_b$, a natural definition of a coupling constant is thus given by
\begin{align}%
\frac{1}{g_I^2}\equiv d_0(N_b)  \frac \alpha {2(1-\alpha)} \,,\qquad\qquad
\frac{1}{g_W^2}=\frac1{g_I^2}\left(1+d_1(N_b) g_I^2+\ldots\right)\,.
\end{align}%

Using the methods and results introduced in Ref.~\cite{Brandt:2014rca}, we determined the coefficient $d_0$
numerically through simulations with both Wilson and induced gauge action for $N_c=2$ in three dimensions. 
These numerical results are shown in Fig.~\ref{Fig:comparison} together with the results from perturbation theory.
We observe surprisingly good agreement even for small values of $N_b$.

\def \errorscale {1}
\def \datacolor {myred}
\def \datacolorB {myblue}
\def \marksize {1.2pt}
\def \errormarksize {2.4pt}

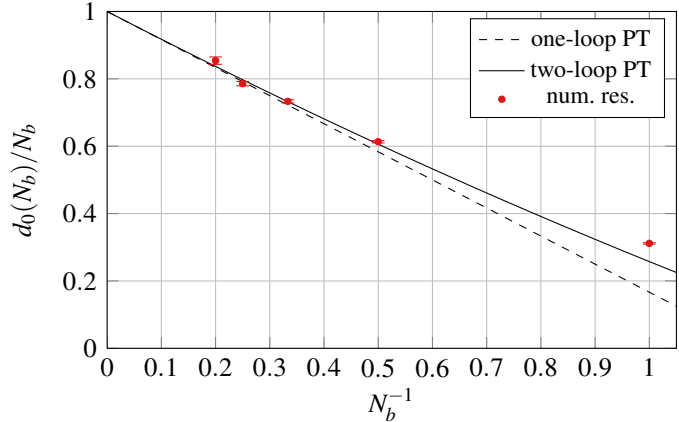
\begin{figure}
\begin{center}
\begin{tikzpicture}
\begin{axis}[
width=0.6\textwidth,
height=0.4\textwidth,
label style={font=\small},
legend style={font=\footnotesize},
tick label style={font=\small},
xlabel ={$N_b^{-1}$},
ylabel ={$d_0(N_b)/N_b$},
x label style={at={(axis description cs:0.5,0.03)},anchor=north},
y label style={at={(axis description cs:0.06,0.5)},anchor=south},
ymin =0,
ymax =1.0,
xmin = 0, 
xmax = 1.05,
minor tick num=0,
xtick={0,0.1,...,1.1},
legend entries={one-loop PT, two-loop PT,  num.~res.},
grid=major
]
\addplot [black,dashed,domain=0:1.1, samples=10]{1-5*x/6};
\addplot [black, domain=0:1.1, samples=200]{1-5*x/6+0.0908283*x*x};
\addplot+[\datacolor,only marks,mark=otimes*,mark size=\marksize,mark options={fill=\datacolor}, error bars/.cd,y dir=both, y explicit, error mark options={rotate=90,\datacolor,mark size=\errormarksize}] 
coordinates {
  (1,0.3115) +- (0,\errorscale*0.002)
  (0.5,0.61325) +- (0,\errorscale*0.0035)
  (0.33333333,0.733167) +- (0,\errorscale*0.0048)
  (0.25,0.78575) +- (0,\errorscale*0.0065)
  (0.2,0.854) +- (0,\errorscale*0.011)
 };
\end{axis}
\end{tikzpicture}
\caption{Perturbative and numerical results for $d_0/N_b$ in $d=3$ with $N_c=2$.}
\label{Fig:comparison}
\end{center}
\end{figure}

\section{Summary and perspectives}

We have shown analytically that the induced $\SU(N_c)$ action exhibits a continuum limit in 2d as $\alpha \to 1$ for fixed $N_b\geq N_c-\frac 54$ 
and derived $N_b\geq N_c-\frac 34$ as a necessary condition for this continuum limit to be in the universality class of YM theory. The equivalence is expected to persist in higher dimensions. For gauge group $\U(N_c)$, the conditions are $N_b\geq N_c-\frac12$ and $N_b\geq N_c+\frac12$, respectively. 
Although perturbation theory for $\alpha \to 1$ is problematic, a relation between the coupling constants can be determined by first taking $N_b\to\infty$ at fixed $\alpha \leq \frac 13$ and analytic continuation to small $1-\alpha$. We observe good agreement with first numerical results for $\SU(2)$ in 3d. 
In the future, we will extend our numerical study to $\SU(3)$ in four dimensions and plan to use the bosonized version of the gauge action for full QCD.

\bibliographystyle{JBJHEP_mod}

\bibliography{inducedQCD,Lambda-Param}

\end{document}